\documentstyle[12pt]{article}

\begin{document}
\title{Stationary Dilatons with Arbitrary Electromagnetic Field}
\author{
Tonatiuh Matos\thanks{Permanent Adress: Dpto. de F\'{\i}sica, CINVESTAV, 
PO. Box 14-740, M\'exico 07000, D.F., M\'exico. E-mail: 
tmatos@fis.cinvestav.mx},\\ Instituto de F\'{\i}sica y Matem\'aticas,\\
Universidad Michoacana de San Nicolas de Hidalgo,\\
PO. Box 2-82, 58040 Morelia, Michoac\'an, M\'exico\\and\\
C\'esar Mora \\
Departamento de F\'\i sica,\\ Centro de Investigaci\'on y de Estudios 
Avanzados del I.P.N.\\PO. Box 14-740, 07000 M\'exico, D.F.}

\date{ \today}
%
\maketitle

\begin{abstract}
We present two new classes of axisymmetric stationary solutions of the 
Einstein-Maxwell-Dilaton equations with coupling constant $\alpha^2=3$. 
Both classes are written in terms of two harmonic maps $\lambda$ and 
$\tau$.  $\lambda$ determines the gravitational potential and $\tau$ the 
electromagnetic one in such a form that we can have an arbitrary 
electromagnetic field. As examples we generate two solutions with mass 
($M$), rotation ($s$) and scalar ($\delta$) parameters, one with 
electric charge ($q$) another one with magnetic dipole ($Q$) parameter. 
The first solution contains the Kerr metric for $q=\delta=0$. 
\end{abstract}


%

\newpage
\section{Introduction}

There exist in the Universe a great amount of astrophysical objects 
possessing gravitational and electromagnetic fields together. Such is 
the case of some planets and stars possessing a magnetic dipole field 
like the Earth or the Sun. The Einstein-Maxwell
 (EM) theory predicts the existence of gravitational objects endowed 
with magnetic dipoles by means of complicate exact solutions 
\cite{manko1,manko2}. EM theory is actually not an unification's theory, 
since here the electromagnetic field appears as energy-momentum tensor, 
there is in fact no explanation for its existence, it is put by hand and 
the electromagnetic field appears like a model. On the other hand, 
five-dimensional (5D) gravity is an alternative theory for understanding 
gravitational and electromagnetic interactions together. In this theory 
the electromagnetic field is a consequence of a more general unified 
field, it is not a model. Nevertheless, it contains an extra field not 
observed in nature, a scalar field called the dilaton. Scalar fields are 
not new in physics, they appear naturally in all the most important 
unified theories, like the Kaluza-Klein \cite{kk} and the super string 
theory, and it is put by hand in unified models like inflation or the 
standard model, but nobody has seen th em.
In \cite{ma45} it is shown that there exists a class of very simple 
static exact solutions of the Einstein-Maxwell-Dilaton field equations 
where the coupling constant $\alpha$ between the scalar field and 
electromagnetism, remains arbitrary. It possesses  
a gravitational and a magnetic dipole field; its four-dimensional (4D) 
metric behaves very similarly than the Schwarzschild solution but 
coupled with a magnetic dipole. They are very simple, but they posses a 
scalar field interaction arising from the compactification from the 5D 
space, which has not yet been observed in astrophysical objects. 
Nevertheless, in \cite{ma45} it is shown that for these solutions the 
dilaton interaction cannot be measured in weak gravitational fields like 
the Sun, even if the Sun would posses one, but it could be probably 
detected in stronger gravitational fields like one of a pulsar. A star 
like the sun is essentially static, but a pulsar is essentially not, 
therefore it is worth to generalize this solution for a rotating body
. One should expect that the scalar field interaction does not modify 
the interaction of test particles for non-compact bodies like the sun, 
even if the rotation is taken into account. But one expects that a 
rotating solution gives more inside in the behavior of rotating compact bodies.

In \cite{ma1}, T. Matos developed a method for generating exact static 
solutions for the 5D Einstein equations with a $G_3$ group of motion, 
putting the solutions in terms of two harmonic maps, $\lambda$ and 
$\tau$. The harmonic map $\lambda$ determines the gravitational field 
and the harmonic map $\tau$ the electromagnetic one. Therefore one can 
choose the electromagnetic field of a monopole, dipole, quadripole, etc. 
In this work we generalize a set of static solutions for $\alpha^2=3$. 
The first class represents an electrically charged static body with mass 
and scalar field parameters, the second one represents a magnetic dipole 
with mass and scalar field parameters \cite{ma45}. We obtain their 
corresponding rotating solution using invariant transformations in the 
potential space. Unfortunately this method can be used only for 
$\alpha^2=3$ or $\alpha^2=0$, only in these two cases the potential 
space is symmetric (see ref. \cite{neu,ma1}). This work is organized as 
follows. In section 2 we introduce the 
potential space formalism  for 5D-gravity. In section 3 we write the 
field equations in chiral form. In section 4 we generalize the first 
class of solutions and in section 5 the second class. We write them in 
terms of two harmonic maps and give some explicit solutions in section 
6. In section 7 we give some conclusions and remarks.

\section{Potential space field equations}

In this section we introduce the potential formalism for 5D-gravity. 
This theory is characterized by the existence of a Killing vector field 
${\bf X}$, which generate the $U(1)$ gauge electromagnetic group.  
Introducing an extra timelike  Killing vector field, Neugebauer 
\cite{neu} introduced the potential formalism in 5D-gravity. The field 
equations are then of stationary fields. This formalism consists in 
defining covariantly five potentials in terms of the two commuting 
Killing vectors ${\bf X}$ and ${
\bf Y}$, ${\bf X}$ being related to the $U(1)$ isometry and ${\bf Y}$ 
being related to stationarity. The five potentials are (see also ref. 
\cite{ma1}) \[
  I^2=\kappa^{4\over 3}=X^\mu X_\mu\ \ \ \ \  f=-I Y^\mu 
Y_\mu+I^{-1}(X^\mu Y_\mu)^2 \]
\[
  \psi=-I^{-2}X^\mu Y_\mu \ \ \ \ \            
  \epsilon_{,\mu}=\epsilon_{\alpha\beta\gamma\delta\mu}X^\alpha 
   Y^\beta Y^{\gamma;\delta}
\]
\begin{equation}
 \chi_{,\mu}=-
  \epsilon_{\alpha\beta\gamma\delta\mu}X^\alpha Y^\beta X^{\gamma;\delta}  
\label{1}
\end{equation}
being $\epsilon_{\alpha\beta\gamma\delta\mu}$ the Levi-Civita 
pseudotensor. In the adapted coordinate system where 
$X=X^A{\partial\over\partial x^A} ={\partial\over\partial x^5}, \quad 
Y=Y^A{\partial\over\partial x^A}=
{\partial\over\partial x^4}$, one finds that 
$\Psi^A=(f,\epsilon,\psi,\chi,\kappa), A=1,...,5$ are the gravitational, 
rotational, electrostatic, magnetostatic and scalar potentials, 
respectively. The five dimensional field equations in terms of the potent
ials (\ref{1}) can be derived from the Lagrangian \cite{neu,ma1}
\begin{equation}
L={\rho\over 2f^2}\left[f_{,i}f^{,i}+ 
(\epsilon_{,i}+\psi\chi_{,i})(\epsilon^{,i}+\psi\chi^{,i})\right]+ 
{\rho\over 2f}\left(\kappa^2\psi_{,i}\psi^{,i}+
{1\over\kappa^2}\chi_{,i}\chi^{,i}\right) +{2\over 3}{\rho\over 
\kappa^2}\kappa_{,i}\kappa^{,i} \label{2}
\end{equation}
(variation with respect to $\Psi^A$). Now we can define a 5D (abstract) 
Riemannian space $V_5$, called the potential space, inspired on the 
Lagrangian (\ref{2}) with metric \begin{equation}
dS^2={1\over 2f^2}\left[df^2+(d\epsilon+\psi\,d\chi)^2\right]+
{1\over2f}\left(\kappa^2\,d\psi^2+{1\over \kappa^2}\,d\chi^2\right)+
{2\over3}\,{d\kappa^2\over \kappa^2}.
\label{3}
\end{equation}
On $V_5$, the five potentials $\Psi^A$ are the local coordinates which 
define a symmetric Riemmanian space, {\it i.e.} the covariant derivative 
of the curvature tensor of $V_5$ with respect to each coordinate, vanishes.

\section{The chiral form of the field equations}

Axisymmetry is represented by the existence of a third space like 
Killing vector field ${\bf Z}$. Then one can choose a coordinate system 
in which the components of the five metric depend only on two 
coordinates. In this case the matrix representation of 
the 5D Einstein field equations in the potential space $V_5$ is \cite{ma2}
\begin{equation}
  (\rho g_{,z}\ g^{-1} )_{,\bar z} + (\rho g_{,\bar z}\ g^{-1})_{,z} = 0  
        \label{4}
\end{equation}
where Weyl's canonical coordinates $\rho$ and $\zeta$ are related by 
$z=\rho + i\zeta$ and its complex conjugate $\bar z$. Matrix $g$ in 
(\ref{4}) is a symmetric matrix, element of the group $SL(3,R)$, $i.e.$
\begin{equation}
g=g^T, \quad g=\bar g, \quad det\,\,g=1
\label{5}
\end{equation}
(T denotes matrix transposition). Using the invariant transformations of 
Lagrangian (\ref{2}), we generate new solutions from a seed one. That 
means that if we have a solution of the field equation $\Psi^A$, an 
invariant transformation (\ref{2}) of the form $\Psi^A \to 
\Psi'^A(\Psi^B)$ will give us a new solution (see ref. \cite{kra}). All 
invariant transformations of (\ref{2}) were found in ref. \cite{ma2}. 
The group of motion of metric (\ref{3}) is $SL(3,R)$, which has 8 
parameters. The invariant transformations of Lagrangian (\ref{2}) can be 
cast into the very simple form  
 \begin{equation}
g=Cg_0C^T 
\label{6}
\end{equation}
where the constant matrix $C$ is also an element of $SL(3,R)$. The 
matrix $g$ can be parametrized in terms of the potentials (\ref{1}) as 
\cite{ma2} \begin{equation}
g={-2\over f\kappa^{2\over 3}}\pmatrix {f^2+\epsilon^2-f\kappa^2\psi^2 & -
\epsilon & -{1\over 2\sqrt2}(\epsilon\chi + f\kappa^2\psi)\cr 
-\epsilon & 1 & {1\over 2\sqrt2}\chi\cr 
-{1\over 2\sqrt2}(\epsilon\chi + f\kappa^2\psi) &
{1\over 2\sqrt2}\chi & {1\over 8}(\chi^2 -\kappa^2f)\cr}.
\label{7}
\end{equation}
The correct choice of the $C$ matrix in (\ref{6}) will generate new 
solutions. We use this method for generating the Belinsky-Ruffini 
solution \cite{beli}.

\section{First class of solutions. Seed solution $\Psi^A=(f, \epsilon, 
0, 0, \kappa).$}

In order to obtain the new solution, we put the potentials $\Psi^A$ in 
terms of the components of the matrix $g$ and its inverse \begin{equation}
g^{-1}=-{1\over 2}{\kappa^{2\over 3}\over f}\pmatrix {1&\epsilon +
\chi\psi&-2\sqrt2\psi\cr\epsilon +\chi\psi&f^2+(\epsilon +\chi\psi)^2-
f\chi^2\kappa^{-2}&2\sqrt2[f\chi\kappa^{-2} - \psi(\epsilon +\chi\psi)]\cr
-2\sqrt2\psi&2\sqrt2[f\chi\kappa^{-2} -\psi(\epsilon 
+\chi\psi)]&-8(f\kappa^{-2} -\psi^2)\cr}, \label{8}
\end{equation}
thus the potentials $\Psi^A$ can be written as
\begin{eqnarray}
\openup1\jot 
    &\kappa^{4\over 3}= {4g_{11}^{-1}\over g_{22}}, 
    \quad f^2 = {1\over g_{11}^{-1}g_{22}}, 
    \quad \chi = 2\sqrt2{g_{23}\over g_{22}},\cr 
    &\psi= {1\over 2\sqrt2}{g_{13}^{-1}\over g_{11}^{-1}}, 
    \quad \epsilon = -{g_{12}\over g_{22}}, 
\label{9}
\end{eqnarray}
where $g_{ij}$ are the components of the matrix $g$, and $g_{ij}^{-1}$ 
the components of the matrix $g^{-1}$. As seed solution we write $f_0, 
\epsilon_0$ and $\kappa_0$ arbitrary and $\psi_0=\chi_0=0$ in matrix 
$g_0$ in (\ref{6}). Using the invariant tran
sformations (\ref{6}) with the $C$ matrix as
\begin{equation}
C=\pmatrix {a&b&c\cr d&e&j\cr i&h&k\cr}=\pmatrix {q&p&t\cr u&v&w\cr 
s&y&z\cr}^{-1},
\label{10}
\end{equation}
we evaluate matrix $g$ using the relations (\ref{9}), obtaining
\begin{eqnarray}
   &\kappa^{4\over 3}={V\over W}\kappa_0^{4\over 3}, \qquad 
   f^2={f_0^2\over VW},\cr
   &\psi= {1\over 2\sqrt2V}\left[(u\epsilon_0 + q)(w\epsilon_0 + t) + 
f_0(wuf_0 - 
   8sz\kappa_0^{-2})\right],\cr
   &\chi={2\sqrt2\over W}\left[(d\epsilon_0- e)(i\epsilon_0 -h) 
+f_0\left(dif_0 -
   {1\over 8}kj\kappa_0^2\right)\right],\cr
  &\epsilon = -{1\over W}\left[(a\epsilon_0 -b)(d\epsilon_0 -e) + 
f_0\left(adf_0 -{1\over 8}cj\kappa_0^2\right)\right],\cr
   \hbox{\rm with}\quad 
&V=(q+u\epsilon_0)^2 + f_0(u^2f_0 - 8s^2\kappa_0^{-2})\cr
   \hbox{\rm and}\quad 
&W= (d\epsilon_0-e)^2 +f_0\left(d^2f_0 - {j^2\kappa_0^2\over 8}\right).\cr
        \label{11}
\end{eqnarray}
Solution (\ref{11}) is endowed with eight new free parameters. This 
transformations (\ref{11}) yields a solution for the potentials $\Psi^A$ 
from an arbitrary seed solution without electromagnetism. If we want to 
obtain an asymptotically flat solution from a same one, we must fix some 
of the eight parameters (see ref \cite{ma2}). In general, one should 
give an explicit seed solution in (\ref{11}) and integrate equations 
(\ref{1}) in order to obtain the corresponding metric in the space-time. 
In ref. \cite
{ma2} we gave an example. Here we provide a general integration for a 
special case of matrix $C$. There exists a class of solutions which can 
be integrated for any seed solution. This class is derived from the matrix 
\begin{equation}
C_M=\pmatrix{q&0&-s\cr 0&1&0\cr -s&0&q\cr},
\label{13}
\end{equation}
with $det\ C_M=q^2-s^2=1$. Substituting $C_M$ into (\ref{10}), the five 
potentials $\Psi^A$ read \begin{eqnarray} 
&f^2={f_0^2\over q^2-8s^2f_0\kappa_0^{-2}}, \quad 
\epsilon=q\epsilon_0, \,\,\,
\chi=2\sqrt2s\epsilon_0,\cr
&\hfill 
\kappa^{4\over3}=\kappa^{4\over 3}_0(q^2 -8s^2f_0\kappa^{-2}_0), \quad 
\psi={1\over 2\sqrt2}{qs(1-8 f_0\kappa_0^{-2})\over q^2-8s^2f_0\kappa_0^{-2}}.
\label{14}
\end{eqnarray} 
Using definitions (\ref{1}) we can integrate (\ref{14}) to obtain the 
space-time metric, we arrive at \[
dS^2={1\over I_0 f_0}e^{2k}dz\,d\bar z+(g_{033}-{8s^2g_{034}^2\over T\ 
I_0^2})\,d\phi^2+2{q g_{034}\over T}d\phi\,dt- {f_0\over I_0 T}dt^2+
\]
\begin{equation}
+I^2\big[-{2\sqrt2 s g_{034}I_0\over T}d\phi
 -{qs\over 2\sqrt2}{1-8f_0\kappa_0^{-2}\over T}dt +dx^5\big]^2,
\label{15}
\end{equation}
where $ T=q^2 -8s^2f_0\kappa_0^{-2}$ (a subindex 0 means seed solution).

Stationary seed solutions of the Einstein equations in terms of harmonic 
maps are well-known (see ref. \cite{kra,pleba}), they are written in 
terms of the  Ernst potential ${\cal E}=f+i\epsilon$. These seed 
solutions can be substituted into (\ref{15}) in 
order to obtain the metric in terms of harmonic maps.
Let us give an example. We start from the Kerr-NUT solution together with 
a $\kappa_0$ potential as a seed solution given by \begin{eqnarray}
   &f_0={\omega-2mr-2lL_+\over \omega}, \quad 
   \epsilon_0=
   {2(mL_+ - lr)\over \omega}, \quad 
   \kappa_0=
   \left(
          {r-m+\sigma \over r-m-\sigma}
   \right)
   ^\delta,\cr
   &L_+=a\cos\theta + l, \qquad L_-=a\cos\theta - l, \qquad 
   \omega=r^2 +(a\cos\theta + l)^2, \cr
         \label{12}
\end{eqnarray}
where $r$ and $\theta$ are the Boyer-Lindquist coordinates; constants 
$a$, $m$ and $l$ are the rotation, mass and NUT parameters, respectively.
Solution (\ref{15}) with seed solution (\ref{12}) is an axisymmetric 
stationary exact solution of 5D gravity. The resulting metric is
\begin{eqnarray}
&dS^2={\omega\over \omega-2mr-2lL_+} \left(
          {r-m-\sigma \over r-m+\sigma}
   \right)^{{2\over 3}\delta}(r^2-2mr+L_+L_-)e^{2k_s}\biggl({dr^2\over 
\Delta}+d\theta^2\biggr)+
\cr
&{1\over D}\left({r-m+\sigma\over r-m-
\sigma}\right)^{{4\over 3}\delta}\times
\biggl\{-(\omega -2m r+2lL_+)dt^2 +
\cr
&
(4qa(mr+l)\sin^2\theta+2l\cos\theta \Delta){\omega-2mr+2lL_+\over 
r^2-2mr+L_+L_-}\,dt\,d\phi +
 \cr
&\biggl[{\omega\over \omega-2mr-2lL_+}
\Delta\sin^2\theta\,D\left({r-m-\sigma\over r-m+\sigma}\right)^{2\delta}- \cr
&
q^2(\omega-2mr-2lL_+)\left({4a\sin^2\theta\,(mr+l)+
2l\cos\theta\Delta\over r^2-2mr+L_+L_-}\right)^2\biggr]\,d\phi^2\biggr\}+
\cr
&\left({r-m+\sigma\over r-m-\sigma}\right)^{{2\over 3}\delta}
{D\over\omega}(A_3\,d\phi+A_4\,dt+dx^5)^2\cr
\label{17}
\end{eqnarray}
where 
\begin{eqnarray}
&A_3= -2\sqrt2\left({r-m+\sigma\over 
r-m-\sigma}\right)^{2\delta}({4a\sin^2\theta\,(mr+l)+
2l\cos\theta\Delta\over r^2-2mr+L_+L_-}){\omega-2mr+2lL_+\over D} , \cr
&A_4=-{qs\over 2\sqrt2}{\omega\left({r-m+\sigma\over r-m-\sigma}
\right)^{2\delta} -8(\omega -2m r+2lL_+)\over D},\cr
&D=\omega\left({r-m+\sigma\over r-m-\sigma}\right)^{2\delta}q^2 
-8s^2(\omega -2m r+2lL_+),\quad \Delta=r^2-2mr+a^2-l^2 \cr
\label{18}
\end{eqnarray}
\[
 e^{2k_s}=6{\left(\sqrt{\rho^2 +(\zeta-m)^2} +\sqrt{\rho^2 
 +(\zeta+m)^2}\right)^2-4m^2 \over 4\sqrt{[\rho^2 + 
(\zeta-m)^2][\rho^2 +(\zeta +m)^2]}}
\]
\begin{equation}
\hbox{\rm with} \quad \rho=\sqrt{r^2-2mr+a^2-l^2}\ \sin\theta  \quad 
\hbox{\rm and} \quad   \zeta=(r-m)\cos\theta, \label{16}
\end{equation}
Here $m, a, q$ and $ \sigma$ are constants related by the  restrictions 
$\sigma^2=m^2+l^2-a^2$, $q^2-s^2=1$.
Metric (\ref{17}) reduces to the Kramer metric \cite{kramer} for 
$C=diag(1,1,1)$ and $l=0$, to the Kerr metric for $C=diag(1,1,1),\ \ 
l=0,\ \ \delta=0$, and to the Belinsky-Ruffini \cite{beli} solution by 
setting $\delta={1\over2},\ \ l=0$.
The study of singularities  of metric (\ref{17}) will be given elsewhere 
\cite{mari}.

\section{Second class of solutions. Seed solution $\Psi^A=(f, 0, 0, 
\chi, \kappa)$.}

From the astrophysical point of view, magnetic dipoles are more 
interesting. Now we start from a magnetized static seed solution, with 
$f,\ \chi$ and $\kappa$ arbitrary and $\epsilon=\psi=0$.
In order to obtain the new solution, we proceed in the same way as in 
the case before. Thus, we find the expressions for the potentials 
\begin{eqnarray}
&\kappa^{4\over3}={4B\over \kappa_0^2\ A}\kappa_0^{4\over 3}\ \ \ \ 
f^2={\kappa_0^{2}\over AB}f_0^2\cr
&\chi=-{1\over \sqrt2\ A}\biggl\{8idf_0^2+8e(h+{1\over 2\sqrt2}k\chi_0)+j[ 
2\sqrt2 h\chi_0+k(\chi_0^2-f_0\kappa_0^{2})]\biggr\}\cr
&\epsilon ={1\over 4\ A}\biggl\{8adf_0^2+8be+2\sqrt2\chi_0(bj+ce) 
+cj(\chi_0^2-f_0\kappa_0^{2})\biggr\}\cr
&\psi= -{1\over 4\sqrt2\ B}\biggl\{tq\kappa_0^{2}+ 
uw(f_0^2\kappa_0^{2}-f_0\chi_0^2)+2\sqrt2f_0\chi_0(ws+uz)-8zsf_0\biggr\}\cr
\end{eqnarray}
where
\[
A=-{1\over 4}\biggl\{8d^2f_0^2+e(8e+4\sqrt2j\chi_0)+ 
j^2(\chi_0^2-f_0\kappa_0^{2})\biggr\} \]
\begin{equation}
B=-{1\over2}\biggl\{q^2\kappa_0^{2}+ 
u^2(f_0^2\kappa_0^{2}-f_0\chi_0^2)+4sf_0(\sqrt2u\chi_0-2s)\biggr\}.
\label{19}
\end{equation}
and again, we integrate solution (\ref{19}) for any seed solution using 
matrix (\ref{12}). So the solution in the potential space takes the form
\begin{eqnarray}
&\kappa^{4\over 3}=\kappa_0^{4\over 3}(q^2-8s^2f_0\kappa_0^{-2})\, , \quad 
&\chi=q\chi_0, \quad \epsilon={1\over 2\sqrt2}s\chi_0\cr
&f={f_0\over \sqrt{q^2-8s^2f_0\kappa_0^{-2}}},\quad 
\psi={1\over 2\sqrt2}
{qs(1-8f_0\kappa_0^{-2})\over q^2-8s^2f_0\kappa_0^{-2}}.\cr
\label{20}
\end{eqnarray}
In the space-time, solution (\ref{20}) can be integrated to obtain
\begin{eqnarray}
\openup 1\jot
&dS^2={1\over I}\biggl\{T^{1\over 2}{e^{k_0}\over f_0}
dz\,d\bar z+\biggl[T^{1\over 2} {\rho^2\over f_0}- 
{8s^2A_{03}^2f_0\over T^{1\over 2}}\biggr]d\phi^2
-{2\sqrt2s A_{03}f_0\over T^{1\over 2}}\,d\phi\,dt-
{f_0\over T^{1\over 2}}\,dt^2\biggr\}
\cr
&\qquad\qquad+I^2\biggl({qA_{03}\over T}\,d\phi-{qs\over 
2\sqrt2}{1-8f_0\kappa_0^{-2}\over T}\,dt+dx^5\biggr)^2.\cr
\label{21}
\end{eqnarray}
where $T=q^2-8s^2f_0\kappa_0^{-2}$ and the electromagnetic four 
potential is given by \begin{equation}
A_3={qA_{03}\over q^2-8s^2f_0\kappa_0^{-2}}\qquad \hbox {and} \qquad 
A_4={-qs(1-8s^2f_0\kappa_0^{-2})\over 2\sqrt2(q^2-8s^2f_0\kappa_0^{-2})},
\label{A3}
\end{equation}
with $A_1=A_2=0$, and the scalar field fulfills the identity $I^3=\kappa^2$. 
We can combine solution (\ref{21}) with the seed solutions in terms of 
harmonic maps, in order to get rotating solutions with arbitrary 
electromagnetic field. This is done in the next section.

\section{Explicit solutions}

In this section we apply the results given in ref. \cite{ma3}, were 
static solutions of 5D gravity were written in terms of two harmonics 
maps $\lambda$ and $\tau$. In \cite{ma3}, many classes of solutions in 
terms of one and two harmonic maps were found.
Here we give only one example. There exist two subclasses of static 
solutions of five-dimensional gravity which are very similar to the 
Schwarzschild space-time (see ref. \cite{ma4}). These subclasses were 
generalized for any dilaton theory in \cite{MNQ}. Here we use their 
five-dimensional version. The static metric reads \cite{ma4,MNQ}
 
\[
ds^2={1\over I}\left\{e^{2(k_s+k_e)}g_{22}^\gamma{dr^2\over 1-{2m\over r}} +
g_{22}^\gamma\ r^2(e^{2(k_s+k_e)}d\theta^2 + \sin^2\theta\ d\phi^2)-
{1-{2m\over r}\over g_{22}^\gamma}\ dt^2\right\}+
\]
\[
I^2(A_3\ d\phi+dx^5)^2
\]
\begin{equation}
A_{3,\zeta}=Q\rho\tau_{,\zeta}\ \ ,\ \ \ \ 
A_{3,\bar {\zeta}}=-Q\rho\tau_{,\bar {\zeta}}\ \ ,\ \ \ \ 
\kappa^2=I^3={h^3e^{\tau_0\tau}\over(1-{2m\over r})
g_{22}^{\beta}}
\label{sol}
\end{equation}
The fuctions $g_{22},\ \ k_s$ and $k_e$ for the subclass a) are 
\[
g_{22}=a_1e^{q_1\tau}+a_2e^{q_2\tau}, 
\]
\[
k_{s,\zeta}={\rho\over 2\alpha^2}(\lambda_{,\zeta}-
\tau_0\tau_{,\zeta})^2,\ \ \ k_{e,\zeta}=-\rho \gamma 
q_1q_2(\tau_\zeta)^2,\ \ \ \tau_0=q_1+q_2, 
\]
and for the subclass b) are
\[
g_{22}=a_1\tau+1, \ \ \ \ k_{s,\zeta}=
{\rho\over 2\alpha^2}(\lambda_{,\zeta})^2,\ \ \ k_e=0,\ \ \ \tau_0=0, 
\]
\noindent
where $\zeta=\rho+i\ z=\sqrt{r^2-2mr}\ \sin\theta+i\ (r-m)\cos\theta$. 
{\bf A}$=A_idx^i,\ i=1,...,4$ is the electromagnetic four potential, $m$ 
the mass parameter, $\gamma={1\over 2}$, $\beta={3\over 2}$; $Q,\ 
a_1+a_2=1,\ q_1$ and $q_2$ are constants subj ected to the restrictions
\[
2\gamma a_1a_2(q_1-q_2)^2+\kappa_0^2Q^2=0
\]
for the subclass a), and 
\[
2\gamma a_1^2-\kappa_0^2Q^2=0
\]
for the subclass b). 
Metric (\ref{sol}) is convenient because we can interpret $m$ as the 
mass parameter, 
$g_{22}$ as the magnetic field contribution to the metric and the 
expression between brackets $\{...\}$ as the four-dimensional space-time 
metric. Hence metric (\ref{sol}) can be interpreted as a 5D magnetized 
Schwarzschild solution. Reference \cite{ma5} 
enlists a set of solutions of the harmonic map equation 
$(\rho\tau_{,\zeta})_{,\bar\zeta}+(\rho\tau_{,\bar\zeta})_{,\zeta}=0$ 
and their corresponding magnetic potential $A_3$. The harmonic map 
$\lambda$ determines the gravitational potential, while the harmonic map 
$\tau$ determines the electromagnetic one. For a magnetic dipole, the 
harmonic map $\tau$ and its corresponding magnetic field read
\begin{equation}
\tau={\cos\theta\over (r-m)^2-m^2\cos^2\theta}\quad 
A_3={Q(r-m)\sin^2\theta\over (r-m)^2-m^2\cos^2\theta}.
\label{26}
\end{equation}
In general, $\tau$ can be chosen in order to obtain monopoles, dipoles, 
quadripoles, etc.

\subsection{Mass and angular momentum}

Now we substitute the values of $g_{033}$, $g_{044}$ and $A_{03}$ from 
metric (\ref{sol}) for an arbitrary electromagnetic field, $i.e.$, we 
substitute (\ref{sol}) into the metric (\ref{21}) to obtain

\begin{eqnarray}
\openup2\jot
&dS^2={1\over I}
\Biggl\{
T^{{1\over 2}}\ e^{2(k_s+k_e)}g_{22}^\gamma({dr^2\over 1-{2m\over r}} +
r^2d\theta^2)+
\cr
&[T^{{1\over 2}}\ g_{22}^\gamma\ r^2 e^{2(k_s+k_e)}\sin^2\theta\ 
-{8s^2\over T^{{1\over 2}}} A_{03}^2{1-{2m\over r}\over g_{22}^\gamma}]d\phi^2
\cr
&-{2\sqrt2\over T^{{1\over 2}}}sA_{03}
{1-{2m\over r}\over g_{22}^\gamma}\,d\phi\,dt- 
{1\over T^{1\over 2}}{1-{2m\over r}\over g_{22}^\gamma}\ dt^2
\Biggr\}+
\cr
&I^2
\biggl({q\over T}A_{03}\,d\phi
-{qs\over T}(1-8{1\over I_0^3}e^{-\tau_0\tau}
(1-{2m\over r})^2g_{22})\,dt +dx^5\biggr)^2, \cr
\label{30}
\end{eqnarray}
where $T=q^2-8s^2{1\over I_0^3}e^{-\tau_0\tau}(1-{2m\over r})^2g_{22}$ 
and $q^2-s^2=1$.
Metric (\ref{30}) is an exact solution of 5D gravity. Setting $s=0$ we 
recover the seed metric (\ref{sol}). For the magnetic field (\ref{26}), 
we can obtain the mass and the electromagnetic parameters. For both 
subclasses a) and b), metric (\ref{30}) cont ains the mass parameter
\begin{equation}
M=m\bigg[1+{8s^2\over h^3}\bigg]=qm.
\label{31}
\end{equation}
and an angular momentum per unit of mass given by

\begin{equation}
a={sQ \over 4M},
\label{32}
\end{equation}
provided that $h=2$.

\subsection{Electromagnetic field components}

For the magnetic dipole, the non-vanishing components of the 
electromagnetic four potential read \begin{equation}
A_3={qQ(r-m)\sin^2\theta\over T[(r-m)^2-m^2\cos^2\theta]},\ \ \ 
A_4=-{qs(1-8{1\over I_0^3}e^{-\tau_0\tau}
(1-{2m\over r})^2g_{22})\over 2\sqrt{2}\,T}
\end{equation}
which correspond to a dipole magnetic field with magnetic charge
\begin{equation}
Q_M=qQ,
\label{37}
\end{equation}
for both subclasses a) and b).

\section{Final Remarks}

We have found two classes of solutions of five-dimensional gravity. The 
first one can be generated from any vacuum solution of the Einstein 
equations, plus a scalar field, the result is a charged solution of 
five-dimensional gravity. In fact, if we write 
the seed metric in Papapetrou form
\begin{equation}
dS^2={1\over I_0}\{{1\over f_0}[e^{2k_0}dz\,d{\bar z}+
\rho^2\,d\phi^2]-f_0(dt+A_0d\phi)^2\}+I_0^2dx^{5^2}, \end{equation}
the resulting metric reads
\begin{equation}
dS^2={1\over I}\{{1\over f}[e^{2k_0}dz\,d{\bar z}+
\rho^2\,d\phi^2]-f(dt+qA_0\,d\phi)^2\}+I^2(A_3\,d\phi+A_4\,dt+dx^5)^2,
\end{equation}
with the functions $f,I,A_3$ and $A_4$ given by
\[
f={f_0\over T^{1\over 2}},\ \ \ \ I=I_0T^{1\over 2},\ \ \ \ \
\]
\[
A_3=-{2\sqrt2sf_0A_0\over T}\ \ \ \ 
A_4=-{qs\over 2\sqrt2}{1-8s^2f_0I_0^{-3}\over T}.
\]

The second class can be generated from a seed static solution of 
five-dimensional gravity which posses magnetic field, $i.e.$
\begin{equation}
dS^2={1\over I_0}\{{1\over f_0}[e^{2k_0}dz\,d{\bar z}+
\rho^2\,d\phi^2]-f_0\,dt^2\}+I_0^2(A_{03}\,d\phi+dx^5)^2,
\end{equation}
giving as result a rotating solution
\begin{equation}
dS^2={1\over I}\{{1\over f}[e^{2k_0}dz\,d{\bar z}+
\rho^2\,d\phi^2]-f(dt-2\sqrt2sA_{03}\,d\phi)^2\}+
I^2(A_{3}\,d\phi+A_4\,dt+dx^5)^2,
\end{equation}
with the functions  $f,I,A_3$ and $A_4$ given by 
\[
f={f_0\over T^{1\over 2}},\ \ \ \ I=I_0T^{1\over 2},\ \ \ \ \
\]
\[
A_3={qA_{03}\over T}\ \ \ \ 
A_4=-{qs\over 2\sqrt2}{1-8s^2f_0I_0^{-3}\over T}.
\]
With the first class we generated some well-known solutions as the 
Belinsky-Rufini space-time. This latter solution is not a Black Hole 
since the scalar field forms a naked singularity because the scalar 
field parameter $\delta$ is fixed. We suspect that 
our solution (\ref{17}) contains Black Holes for different values of 
$\delta$. At last the Kerr Black Hole is contained here for 
$l=\delta=0$, but we do not know at the moment if other Black Holes are 
contained in metric (\ref{17}). The behavior of this metric is in some 
sense similar to the Kerr-Newman Black Hole. It contains a magnetic 
dipole moment which vanishes for the rotation parameter $a=0$. This 
means that the magnetic dipole is provoked by induction from the 
electric field. (\ref{17}) represents
a rotating electrically charged mass, with an induced magnetic dipole. 
The second class of solutions is quit different. If the rotation 
parameter $s$ vanishes, the solution becomes static and the magnetic 
field is just of a magnetically charged sphere, but the electric charge 
vanishes as well. This means that here the electric charge is induced by 
the rotation of the magnetic dipole. This solution represents a rotating 
mass with a magnetic dipole charge. Quit surprising is that the magnetic 
field does not alter during the rotation, which can be seen from 
relations (\ref{20}). Nevertheless, there is no way to conserve the 
rotation dropping out the magnetic field, because if $Q=0$, then the 
rotations parameter $a$ vanishes as well. 

\section{Acknowledgments.}

We want to thank Hernando Quevedo and Dario Nu\~nez for many helpful 
discussions. This work was partially supported by CONACyT-M\'exico.

\end{document}